\documentclass[12pt]{article}
\usepackage{graphicx}

\font\grande=cmr9.5 scaled \magstep4
\font\medio=cmr9.5 scaled \magstep2
\outer\def\beginsection#1\par{\medbreak\bigskip
      \message{#1}\leftline{\bf#1}\nobreak\medskip
\vskip-\parskip
      \noindent}

\begin{document}
\bibliographystyle {unsrt}

\titlepage

\begin{flushright}
CERN-PH-TH/2008-233
\end{flushright}

\vspace{15mm}
\begin{center}
{\grande Birefringence, CMB polarization}\\
\vspace{3mm}
{\grande and magnetized B-mode }\\
\vskip1.cm
 
Massimo Giovannini$^{a,b}$ and Kerstin E. Kunze$^{a,c}$
 
\vskip1cm
{\sl $^a$  Department of Physics, Theory Division, CERN, 1211 Geneva 23, Switzerland}
\vskip 0.2cm
{\sl $^b$ INFN, Section of Milan-Bicocca, 20126 Milan, Italy}
\vskip 0.2cm
{\sl $^c$  Departamento de F\'\i sica Fundamental, \\
 Universidad de Salamanca, Plaza de la Merced s/n, E-37008 Salamanca, Spain}
 
\vspace{6mm}
\end{center}

\vskip 2cm
\centerline{\medio  Abstract}
Even in the absence of a sizable tensor contribution,  a B-mode polarization can be generated because of the competition between a pseudo-scalar background and 
 pre-decoupling magnetic fields. By investigating the dispersion relations of a magnetoactive plasma supplemented by a pseudo-scalar interaction,
the total B-mode polarization is shown to depend  not only upon the plasma and Larmor frequencies  but also on the pseudo-scalar rotation rate. 
 If the (angular) frequency channels of a given experiment are larger than the pseudo-scalar rotation rate, the only possible source of (frequency dependent) B-mode autocorrelations must be attributed to Faraday rotation.  In the opposite case the pseudo-scalar contribution dominates and the total rate becomes, in practice,  frequency-independent. The B-mode cross-correlations can be used, under certain conditions, to break the degeneracy 
 by disentangling  the two birefringent contributions. 
 \noindent

\vspace{5mm}

\vfill
\newpage
In the $\Lambda$CDM paradigm\footnote{The $\Lambda$ refers to the dark energy component (assumed to be in the form 
of a putative cosmological constant). The CDM refers to the 
(cold) dark matter contribution. In what follows 
the cosmological parameters will be fixed to the best 
fit of the WMAP-5yr data alone, i.e. $(\Omega_{\mathrm{b}0},\, \Omega_{\mathrm{c}0}, \,\Omega_{\mathrm{\Lambda}}, \,h_{0}, \,n_{\mathrm{s}},\,\epsilon)= (0.0441,\, 0.214,\, 0.742,\, 0.719,\, 0.963,\,0.087)$. In the latter 
string of parameters $\Omega_{X}$ denotes the critical fraction of a given species, $h_{0}$ fixes the present 
value of the Hubble rate; $n_{\mathrm{s}}$ is the spectral index of curvature perturbations and $\epsilon$ is the reionization optical depth.} a potential 
candidate for the B-mode polarization are the tensor modes of the geometry 
inducing a frequency-independent polarization of the Cosmic 
Microwave Background (CMB in what follows).  By {\em frequency-independent 
signal} we mean that different observational channels measure angular power spectra with the same amplitude.
In the opposite case the angular power spectra  will effectively depend upon the (angular) frequency of observation.
The WMAP 5-yr data \cite{WMAP5yr1} constrain 
the presence of a B-mode and, indirectly, $r_{\mathrm{T}}$, i.e. the ratio of the tensor power spectrum over the scalar power spectrum \cite{WMAP5yr1}. 
A  further (frequency independent) source of B-mode 
polarization is cosmic shear (see e.g. \cite{hirata}). Diverse data sets (such as the ones of Quad and Capmap \cite{poldata}) impose concurrent limits on the B-mode polarization. Forthcoming experiments are expected to improve the present status of the observations by reaching into the region $r_{\mathrm{T}} < 0.2$. 

The only frequency-dependent signal investigated so far is provided by the Faraday effect which is a distinctive feature of magnetized plasmas in different contexts \cite{faraday}.  Large-scale magnetic fields present prior to the equality time are known to impact both on the temperature autocorrelations as well as on the polarization observables \cite{mg1}.  It has been recently shown, within a dedicated numerical approach \cite{gk1}, that the Faraday rotation signal induced by a pre-decoupling magnetic field can overwhelm the B-mode polarization induced by the standard tensor contribution  \cite{gk2}. 

The B-mode autocorrelations might not be always sufficient  to infer the presence of a pre-equality magnetic field\footnote{Following the established terminology the B-mode autocorrelations are denoted by BB. With similar 
notation we will talk about the TT, TE, EE angular power spectra meaning, respectively, the autocorrelations of the temperature, the autocorrelations of the E-mode and their mutual cross-correlations.}.  In short, the idea is the following. Consider a set-up where the pre-equality plasma is birefringent 
because of the concurrent presence of a pseudo-scalar background field, be it $\sigma$, and of a large-scale 
magnetic field. Absent any pseudo-scalar background, the rotation rate would 
scale with the square of the wavelength \cite{faraday} of the observational channel. In the presence of a pseudo-scalar field the dispersion relations can be generalized and the total rotation rate 
will have, both, a magnetic and a pseudo-scalar contribution.  The purpose of this paper is to compute 
the B-mode polarization generated by the competition 
of the two aforementioned effects and to scrutinize 
if (and when) the two effects can be, at least partially, disentangled.
 The essentials of the problem at hand are usefully introduced in terms of the electromagnetic part of the action
\begin{equation}
S_{\mathrm{em}} = - \frac{1}{16\pi} \int d^{4} x \sqrt{-g} \biggl[ F_{\mu\nu} F^{\mu\nu} - \frac{\beta\sigma}{M} 
F_{\mu\nu} \tilde{F}^{\mu\nu} + 16\pi j^{\nu} A_{\nu}\biggr]  
\label{EQ1}
\end{equation}
where $g =\mathrm{det}g_{\mu\nu}$ and $g_{\mu\nu} = a^2(\tau) \eta_{\mu\nu}$; $F_{\mu\nu}$
is the electromagnetic field strength;  $\tilde{F}^{\mu\nu} = \epsilon^{\mu\nu\rho\sigma} F_{\rho\sigma}/(2\sqrt{-g})$ 
is the dual field strength in curved space-times. In Eq. 
 (\ref{EQ1}) $\beta$ is a coupling constant and $M$ a typical mass scale which 
 may take specific values, for instance, in a given scenario \cite{mg2}. In Eq. (\ref{EQ1}) $j^{\nu}$ 
denotes the electromagnetic current which can be specified in terms of the charge carries (i.e. electrons and ions)  as $j^{\nu} = j^{\nu}_{\mathrm{i}} + j_{\mathrm{e}}^{\nu} = e (\tilde{n}_{\mathrm{i}}\,u^{\nu}_{\mathrm{i}} - \tilde{n}_{\mathrm{e}} \,u^{\nu}_{\mathrm{e}})$ (recall that, for both species, 
$g_{\mu\nu} u_{\mathrm{e,\,i}}^{\mu} u_{\mathrm{e,\,i}}^{\nu} =1$). The relevant set of equations can then be written, for brevity, in their covariant form and they are 
\begin{eqnarray}
&&\nabla_{\mu} F^{\mu\nu} = 4\pi j^{\nu} + \frac{\beta}{M} \nabla_{\mu} \sigma \tilde{F}^{\mu\nu}, \qquad 
\nabla_{\mu} \tilde{F}^{\mu\nu} =0,
\label{EQ3}\\
&& \nabla_{\mu} 
(T^{\mu\nu}_{\mathrm{e}} + T^{\mu\nu}_{\mathrm{i}} + T^{\mu\nu}_{\mathrm{EM}}) =0,\qquad T^{\mu\nu}_{\mathrm{e,i}} = \rho_{\mathrm{e,i}}\,u^{\mu}_{\mathrm{e,i}}\, u^{\nu}_{\mathrm{e,i}},
\label{EQ4}
\end{eqnarray}
where $T^{\mu\nu}_{\mathrm{EM}}$ is the energy-momentum tensor of the electromagnetic field and $\rho_{\mathrm{e,i}}$ denote the energy density 
of electrons and ions. 
Note that $\nabla_{\mu}$ (i.e. the covariant derivative 
associated with the space-time geometry) does not 
only depend upon the scale factor but also upon the 
inhomogeneities.  
Equation (\ref{EQ4}) summarizes schematically  the evolution equations of charged species whose 
governing equations can be more appropriately 
derived from the Vlasov-Landau equations in curved space (or from their lowest moments).
While electrons and ions are coupled through Coulomb scattering, the electron-ion fluid is coupled to the photon background. The plasma contains a large-scale magnetic field whose 
Fourier amplitudes satisfy
\begin{equation}
\langle B_{i}(\vec{k},\tau) B_{j}(\vec{p},\tau) \rangle = \frac{2 \pi^2}{k^3} P_{\mathrm{B}}(k) P_{ij}(k) \delta^{(3)}(\vec{k}+ \vec{p}), \qquad P_{\mathrm{B}} = A_{\mathrm{B}} \biggl(\frac{k}{k_{\mathrm{L}}}\biggr)^{n_{\mathrm{B}} -1}
\label{EQ6}
\end{equation}
 where $k^2 P_{ij}(k) = ( k^2\delta_{ij} - k_{i} k_{j})$; 
 $A_{\mathrm{B}}$ is the amplitude of the magnetic power spectrum at the (comoving) magnetic pivot scale $k_{\mathrm{L}}$ (equal to $1\,\mathrm{Mpc}^{-1}$ in the forthcoming numerical examples). The magnetic field is inhomogeneous 
over typical length-scales which are of the order of the Hubble radius $r_{\mathrm{H}}$.
The Larmor radius of the electrons, on the contrary, is $r_{\mathrm{L}} \simeq {\mathcal O}(v_{\mathrm{th}}/\overline{\omega}_{\mathrm{Be}})\ll r_{\mathrm{H}}$ where $\overline{\omega}_{\mathrm{Be}}$ is the (comoving) Larmor frequency and 
$v_{\mathrm{th}} \simeq \sqrt{T_{\mathrm{e}}/m_{\mathrm{e}}}$ is the thermal velocity of the electrons. 
 For a comoving field strength ${\mathcal O}(\mathrm{nG})$ the Larmor radius is roughly eight orders of magnitude smaller than the Hubble radius. The guiding centre approximation (originally due to Alfv\'en \cite{alfven}) can be then applied. The charged particles orbiting around the magnetic field lines will see, in practice, a constant 
field up to drift corrections (going as $[T_{\mathrm{e}} (\vec{B} \times\vec{\nabla})B]/(e B^3)$ where $B = |\vec{B}|$) and curvature corrections (going as $[T_{\mathrm{e}}\vec{B}\times (\vec{B}\cdot\vec{\nabla})\vec{B}]/(eB^4)$ ) which are, however, negligible when the  scale of (spatial) variation of the magnetic background is much larger than the gyration radius of the charge carriers. 
As discussed in \cite{mg2} the dispersion relations can be derived by studying the propagation  of the 
electromagnetic waves in a magnetoactive plasma at finite density. By writing 
Eqs. (\ref{EQ3})--(\ref{EQ4}) in their explicit form the compatibility of the system can be ensured 
if $\mathrm{det}{\mathcal A}_{ij} =0$ where ${\mathcal A}_{ij}$ is given by 
\begin{equation}
{\mathcal A}_{ij} = k^2 (\delta_{ij} - \hat{k}_{i} \hat{k}_{j}) - \overline{\omega}^2 \overline{\epsilon}_{ij}(\overline{\omega},\alpha) + i \frac{\beta}{M} \sigma' \epsilon_{m ij} k^{m}.
\label{EQ8}
\end{equation}
In Eq. (\ref{EQ8}) $\overline{\epsilon}_{ij}(\overline{\omega},\alpha)$ is the dielectric tensor and $\sigma'$ denotes a derivation with 
respect to the conformal time coordinate $\tau$. By requiring that 
$\mathrm{det}{\mathcal A}_{ij} =0$ and by setting the comoving wavenumbers in such a way that 
$k_{x}=0$ and $k_{y} = k \sin{\vartheta}$ and $k_{z} = k \cos{\vartheta}$ the standard form 
of the Appleton-Hartree equation can be easily recovered and it is 
\cite{mg2} 
\begin{eqnarray}
&&\sin^2{\vartheta}\biggl\{ \biggl( \frac{1}{\epsilon_{\parallel}} - \frac{1}{n^2} 
\biggr) \biggl[ \frac{1}{n^2} - \frac{1}{2} \biggl( \frac{1}{\epsilon_{-}}
+ \frac{1}{\epsilon_{+}}\biggr) \biggr] + 
\frac{\overline{\omega}_{\sigma}^2}{2 n^2\overline{\omega}^2}\biggl(\frac{1}{\epsilon_{-}}
+ \frac{1}{\epsilon_{ +}} \biggr) 
\nonumber\\
&-&
\cos^2{\vartheta}\biggl[\biggl( \frac{1}{n^2} - \frac{1}{\epsilon_{-}}
\biggr)\biggl( \frac{1}{n^2} - \frac{1}{\epsilon_{+}}\biggr) - 
\frac{\overline{\omega}_{\sigma}^2}{n^2 \omega^2\epsilon_{+} \epsilon_{-}}\biggr]
+ \frac{\overline{\omega}_{\sigma}}{n^3 \overline{\omega}}\biggl( \frac{1}{\epsilon_{-}} - 
\frac{1}{\epsilon_{+}}\biggr) \cos{\vartheta} =0,
\label{EQ9}
\end{eqnarray}
where $\alpha = i {\mathcal H}/\overline{\omega}$, ${\mathcal H} = a'/a$ and $n = k/\overline{\omega}$ is the refractive index. Denoting with $\overline{\omega}_{\mathrm{p\,e,i}}$  the comoving plasma frequencies for electrons and ions and with
 $\overline{\omega}_{\mathrm{B\,e,i}}$, the corresponding gyration frequencies $\epsilon_{\pm}(\overline{\omega},\alpha)$ and 
$\epsilon_{\parallel}(\overline{\omega},\alpha)$ are  
\begin{eqnarray}
&& \epsilon_{\pm}(\overline{\omega},\alpha) = \epsilon_{1} \pm \epsilon_{2} = 1 - \frac{ \overline{\omega}^2_{\rm p\,i}}{\overline{\omega}[ \overline{\omega} (\alpha + 1) \mp \overline{\omega}_{\rm B\,i}]} 
-\frac{ \overline{\omega}^2_{\rm p\,e}}{\overline{\omega}[ \overline{\omega} (\alpha + 1) 
\pm \overline{\omega}_{\rm B\,e}]},
\label{EQ10}\\
&& \epsilon_{\parallel}(\overline{\omega},\alpha) = 1 - \frac{ \overline{\omega}_{\rm pi}^2}{\overline{\omega}^2 (1 +\alpha)} - \frac{ \overline{\omega}_{\rm pe}^2}{\overline{\omega}^2 (1 +\alpha)}.
\label{EQ11}
\end{eqnarray}
The frequency $\overline{\omega}_{\sigma} = \beta \sigma' /M$ measures 
the rate of variation of the polarization because of the presence of the pseudo-scalar background. If $\sigma$ is not homogeneous 
the dispersion relations will have a different form. According to Eq. (\ref{EQ9}) the refractive indices for 
electromagnetic propagation along the magnetic field (i.e. $\vartheta=0$)
can be deduced from 
\begin{equation}
 \biggl( n^2 -  \frac{\omega_{\sigma}}{\omega}n - \epsilon_{+}\biggr) 
\biggl(n^2 + \frac{\omega_{\sigma}}{\omega} n - \epsilon_{-}\biggr) =0,
\qquad \vartheta =0.
\label{EQ12}
\end{equation}
In the orthogonal direction the dispersion relations lead to the so-called ordinary and extraordinary plasma 
waves which are, however, not excited because of the minute values of the (comoving) plasma and Larmor 
frequencies for the electrons:
\begin{equation}
\overline{\omega}_{\mathrm{Be}} =0.01759 \biggl(\frac{\hat{n}\cdot\vec{B}}{\mathrm{nG}}\biggr) \,\, \mathrm{Hz},
\qquad  \overline{\omega}_{\mathrm{pe}} = 28.5 \,  \, \biggl( \frac{h_{0}^2 \Omega_{\mathrm{b}0}}{0.02273}\biggr)^{1/2} \,\, \mathrm{Hz}.
\label{EQ13}
\end{equation} 
Indeed $\overline{\omega}_{\mathrm{max}} >  \overline{\omega}_{\mathrm{pe}} \gg \overline{\omega}_{\mathrm{Be}}$ 
where $\overline{\omega}_{\mathrm{max}} = 2\pi \overline{\nu}_{\mathrm{max}}$ and 
$\overline{\nu}_{\mathrm{max}} = 222.617\,\, \mathrm{GHz}$ corresponds to the maximum of the CMB spectral energy density.  The rotation rate experienced 
by the linearly polarized CMB travelling parallel to the magnetic 
field direction is 
\begin{equation}
{\mathcal F}(\hat{n})= \frac{d\Phi}{d\tau} = \frac{\overline{\omega}}{2c} \biggl[ \frac{\overline{\omega}_{\sigma}}{\overline{\omega}}
+ \sqrt{\frac{1}{4}  \biggl(\frac{\overline{\omega}_{\sigma}}{\overline{\omega}}\biggr)^2 
+ \epsilon_{+}(\overline{\omega},\alpha)} -\sqrt{\frac{1}{4}  \biggl(\frac{\overline{\omega}_{\sigma}}{\overline{\omega}}\biggr)^2 + \epsilon_{-}(\overline{\omega},\alpha)}\biggr].
\label{EQ14}
\end{equation}
Since $\omega_{\sigma} \propto \sigma'$ the contribution to the rate can be, in principle, either positive or negative. This will affect the sign of the cross-correlations 
(e.g. the TB and EB angular power spectra).  If $\overline{\omega} < \overline{\omega}_{\sigma}$, the shift in the polarization plane of the CMB will essentially 
be independent upon the channel of observation \footnote{Different experiments 
are characterized by different channels of observations. For instance Quad \cite{poldata} employs two series of 
bolometers located, respectively, at 100 GHz and at 150 GHz.}.
In the opposite case (i.e. $\overline{\omega}> \overline{\omega}_{\sigma}$) 
the magnetized and the pseudo-scalar contribution 
concur in determining the total amount of rotation and, ultimately, the 
various polarization observables, i.e., according to Eq. (\ref{EQ14}) $\psi(\hat{n},\tau) = \Phi_{\sigma}(\tau) + \Phi_{\mathrm{Faraday}}(\hat{n},\tau)$.
\begin{figure}[!ht]
\centering
\includegraphics[height=5.5cm]{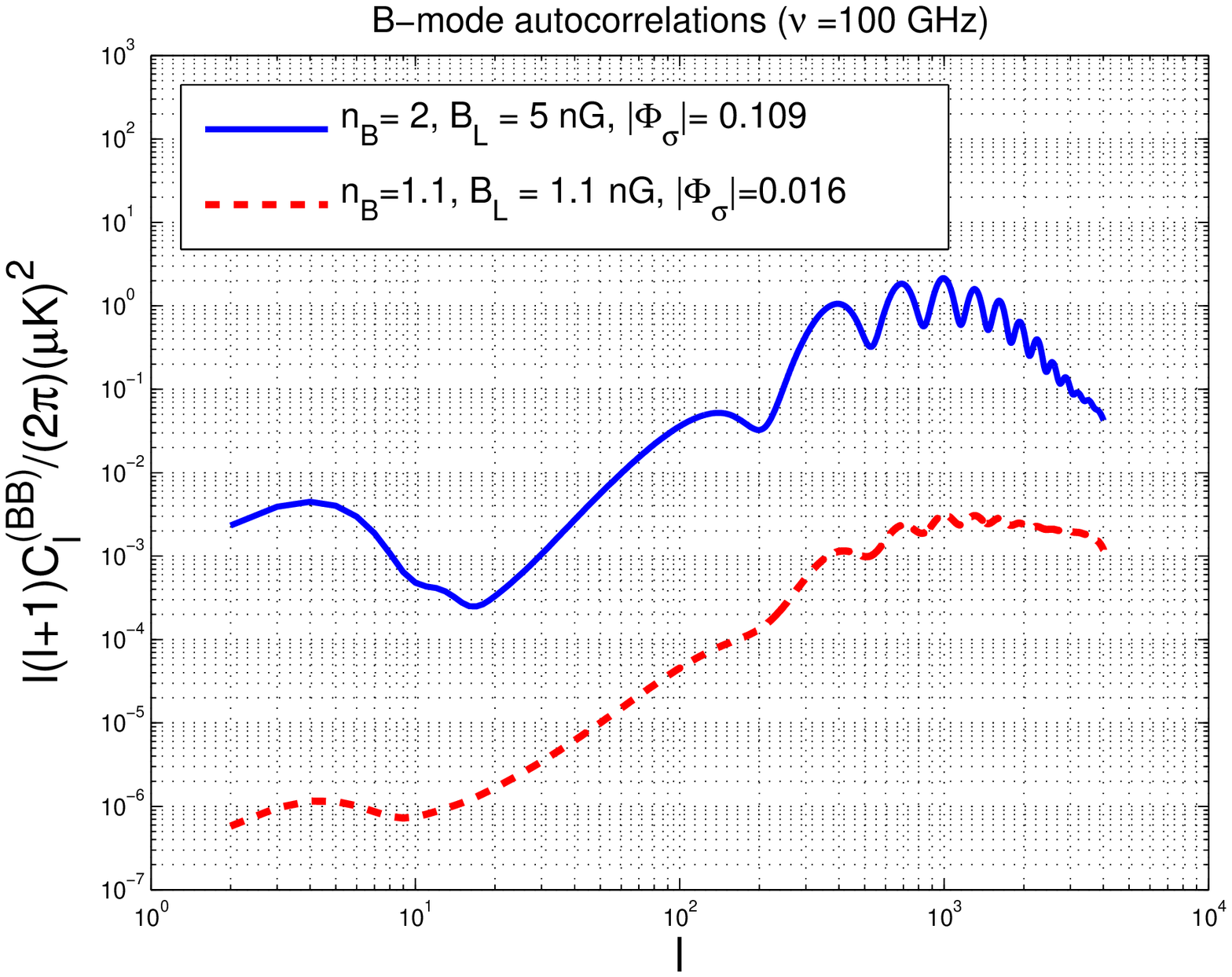}
\includegraphics[height=5.6cm]{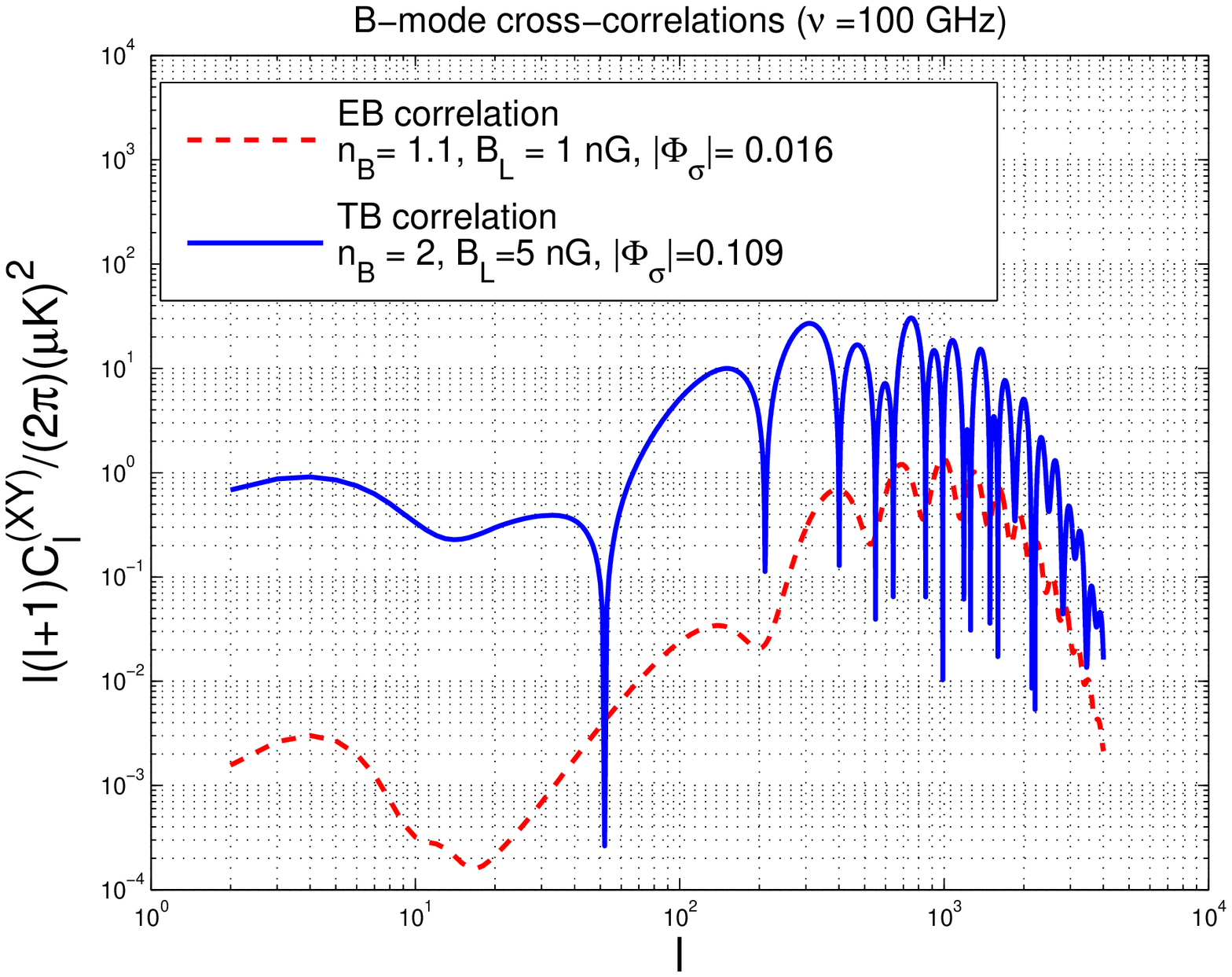}
\caption[a]{The angular power spectra of the B-mode autocorrelations (plot at the left) and the absolute values of  the cross-correlations (plot at the right) are reported in the case when the pseudo-scalar background 
and the magnetized background are simultaneously present. The $\Lambda$CDM parameters have been chosen in accordance with the best fit 
to the WMAP5-yr data alone \cite{WMAP5yr1}.}
\label{F1}      
\end{figure}
The B-mode polarization induced by the magnetoactive plasma in the presence 
of a pseudo-scalar background can be computed by means of 
 an iterative approach which generalizes the calculation of \cite{gk2}. 
 Since $\Delta_{\pm}(\hat{n},\tau) = \Delta_{\mathrm{Q}}(\hat{n},\tau) \pm i 
 \,\Delta_{\mathrm{U}}(\hat{n}\tau)$, it transforms as a spin $\pm$2 for rotations 
around a plane orthogonal to the direction of propagation of the radiation. The three-dimensional rotations and the rotations on the tangent plane of the sphere at a given point combine to give a $O(4)$ symmetry group \cite{sud}. Generalized ladder operators raising (or lowering) the spin weight of a given function can then be defined as \cite{sud,zalda}:
\begin{equation}
K_{\pm}^{\mathrm{s}}(\hat{n})= - (\sin{\vartheta})^{\pm\mathrm{s}}\biggl[ \partial_{\vartheta} \pm
\frac{i}{\sin{\vartheta}} \partial_{\varphi}\biggr] (\sin{\vartheta})^{\mp\mathrm{s}},\qquad \hat{n} = (\vartheta,\, \varphi).
\label{EQ15}
\end{equation}
In real space the E-mode and the B-mode polarization will have spin weight $s=0$:
\begin{eqnarray}
&& \Delta_{\mathrm{E}}(\hat{n},\tau) = - \frac{1}{2} \{ K_{-}^{(1)}(\hat{n})[K_{-}^{(2)}(\hat{n})
\Delta_{+}(\hat{n},\tau)] +  K_{+}^{(-1)}(\hat{n})[K_{+}^{(-2)}(\hat{n}) \Delta_{-}(\hat{n},\tau)]\},
\label{eq13}\\
&&  \Delta_{\mathrm{B}}(\hat{n},\tau) =  \frac{i}{2} \{ K_{-}^{(1)}(\hat{n})[K_{-}^{(2)}(\hat{n})
\Delta_{+}(\hat{n},\tau)] -  K_{+}^{(-1)}(\hat{n})[K_{+}^{(-2)}(\hat{n})\Delta_{-}(\hat{n},\tau)]\}. 
\label{eq14}
\end{eqnarray}
The heat transfer 
equation will then contain, in Fourier space, a convolution. Since the polarization is generated rather close to last scattering the iterative procedure of \cite{gk2} implies that, to zeroth order in the rotation rate, the polarization is given, in real space, as 
\begin{eqnarray}
&& \Delta_{\mathrm{P}}(\hat{n},\tau) = \frac{1}{(2\pi)^{3/2}}\int d^{3} k \,\,\Delta_{\mathrm{P}}(k,\mu,\tau_{0}), \qquad S_{\mathrm{P}} = \Delta_{\mathrm{P}0} + 
\Delta_{\mathrm{P}2} + \Delta_{\mathrm{I}2}
\nonumber\\
&& \Delta_{\mathrm{P}}(k,\mu,\tau_{0}) = \frac{3}{4}(1-\mu^2) \int_{0}^{\tau_{0}}{\mathcal K}(\tau) S_{\mathrm{P}}(k,\tau) e^{- i k\mu (\tau- \tau_{0})} d\tau,
\label{eq15}
\end{eqnarray}
where ${\mathcal K}(\tau)$ is the visibility function. In terms of $\Delta_{\mathrm{P}}(\hat{n},\tau)$, Eqs. (\ref{eq13}) and (\ref{eq14}) imply
\begin{eqnarray}
&&\Delta_{\mathrm{E}}(\hat{n}, \tau) = -  \partial_{\mu}^{2} \{\cos{[2 \psi(\hat{n},\tau)]}( 1 - \mu^2) 
\Delta_{\mathrm{P}}(\hat{n},\tau)\},
\nonumber\\
&& \Delta_{\mathrm{B}}(\hat{n}, \tau) = \partial_{\mu}^{2}\{ \sin{[2\psi(\hat{n},\tau)]}( 1 - \mu^2) \Delta_{\mathrm{P}}(\hat{n},\tau)\},
\label{eq16}
\end{eqnarray}
where, as usual, $\mu = \cos{\vartheta}$ and $\partial_{\mu}$ denotes a derivation with respect to $\cos{\vartheta}$.
In the absence of the (inhomogeneous) magnetized contribution Eq. (\ref{eq16}) leads to the expressions 
customarily used in standard analyses, i.e. for instance
\begin{equation}
 C_{\ell}^{(\mathrm{BB})} = \sin^2{2\psi} \, \overline{C}_{\ell}^{(\mathrm{EE})},\qquad C_{\ell}^{(\mathrm{EB})} = \frac{1}{2}\sin{4\psi} \overline{C}_{\ell}^{(\mathrm{EE})},\, \qquad C_{\ell}^{(\mathrm{TB})} = \sin{2\psi} \, \overline{C}_{\ell}^{(\mathrm{TE})},
\label{eq18}
\end{equation}
where $\psi(\hat{n},\tau) = \Phi_{\sigma}(\tau) \simeq \Phi_{\sigma}(\tau_{0})$ is fully homogeneous and where the $\overline{C}_{\ell}^{(\mathrm{EE})}$ and  $\overline{C}_{\ell}^{(\mathrm{TE})}$ are computed from $\Delta_{\mathrm{P}}(\hat{n},\tau)$. If also the magnetized 
contribution is taken into account the situation changes both qualitatively and quantitatively. This aspect can be 
understood from Eq. (\ref{eq16}). In the limit of small rotation rate  
\begin{equation}
\Delta_{\mathrm{E}}(\hat{n}, \tau) = -  \partial_{\mu}^{2} [( 1 - \mu^2) 
\Delta_{\mathrm{P}}(\hat{n},\tau)],\qquad \Delta_{\mathrm{B}}(\hat{n}, \tau) = 2\, \partial_{\mu}^{2} [ \psi(\hat{n},\tau)( 1 - \mu^2) \Delta_{\mathrm{P}}(\hat{n},\tau)],
\label{eq19}
\end{equation}
where $\psi(\hat{n},\tau)$ and $\Delta_{\mathrm{P}}(\hat{n},\tau)$ depend upon the same 
point on the microwave sky. Eq. (\ref{eq19}) holds in real space. In Fourier space the B-mode would be a convolution. In multipole space the angular power spectra inherit a peculiar form which contains a double sum involving also a known Wigner coefficient arising from the integral of three Legendre polynomials \cite{gk2}. 

In Fig. \ref{F1} (plot at the left) the B-mode autocorrelations are reported in the case when the magnetized background competes with the pseudo-scalar background.  The B-mode signal 
is frequency dependent. In the plot at the right the cross-correlations of the B-mode with the 
other CMB observables are illustrated. The observational 
frequency channel has been taken, for illustration, $\overline{\nu} = 100 \mathrm{GHz}$. In Fig. \ref{F1} 
$B_{\mathrm{L}}$ denotes the magnetic field intensity regularized over a the pivot length-scale $k_{\mathrm{L}}^{-1} \simeq 
\mathrm{Mpc}$. If the pseudo-scalar contribution is totally subleading the $C_{\ell}^{\mathrm{BB}}$ angular power spectrum 
will diminish with the frequency as $\overline{\nu}^{-4}$. Still the cross-correlations of the B-mode with 
the temperature and the E-mode polarization (i.e. $C_{\ell}^{\mathrm{TB}}$ and $C_{\ell}^{\mathrm{EB}}$) are non-vanishing. 
This aspect is illustrated in Fig. \ref{F1} (plot at the right) where $C_{\ell}^{\mathrm{TB}}$ and $C_{\ell}^{\mathrm{EB}}$ are reported. 
The numerical calculation 
leading to the results reported in Fig. \ref{F1}   has been performed 
by including the magnetic field in the initial conditions and at every step 
of the Einstein-Boltzmann hierarchy and for the initial conditions 
corresponding to the magnetized adiabatic mode. 

In summary, 
 if observations  point towards a frequency dependence 
of the B-mode polarization Faraday rotation is probably the only candidate. The cross-correlations 
(i.e. the EB and TB spectra) are expected to vanish in the case of a stochastic magnetic field leaving unbroken spatial isotropy \cite{mg1}.  If they are observed this means that 
the rotation rate is quasi-homogeneous and a pseudo-scalar 
background field may be around. In the latter case, if 
$\overline{\omega}> \overline{\omega}_{\sigma}$ the B-mode will 
scale with frequency as dictated by the Faraday effect while the 
EB and TB correlations will allow to measure independently $\omega_{\sigma}$. In the opposite case (i.e. $\overline{\omega} < \overline{\omega}_{\sigma}$) the frequency dependence 
induced by the Faraday effect is overwhelmed by the (homogeneous) pseudo-scalar rate. In this second case  
the effects of the primordial magnetic fields will be imprinted on the  EE and TT angular power spectra \cite{mg1,gk1} but the B-mode autocorrelations will be independent of the frequency. This 
demonstrate that the analysis of the B-mode autocorrelation is 
necessary but might insufficient, if taken individually,  to infer the 
existence of pre-decoupling magnetic fields.

K.E.K. is supported by the ``Ram\'on y Cajal''  program and by the grants FPA2005-04823, FIS2006-05319 and CSD2007-00042 of the Spanish Science Ministry.

\end{document}